\documentclass[10pt,conference,letter]{IEEEtran}
\usepackage[utf8]{inputenc}

\usepackage{cite}
\usepackage{amsmath,amssymb,amsfonts}
\usepackage{algorithmic}
\usepackage{textcomp}

\usepackage{setspace}

\def\smallerspacecaption{\vspace{-2mm}}

\usepackage{floatrow}
\usepackage[caption=false,font=footnotesize]{subfig}

\usepackage[table, dvipsnames]{xcolor}  
\usepackage{fancyhdr}
\usepackage{graphicx}
\usepackage{url}
\usepackage{verbatim}
\usepackage{multirow}
\usepackage{hhline}
\usepackage[us]{datetime}
\usepackage{paralist}
\usepackage{color}
\usepackage{soul}
\usepackage[implicit=false,hidelinks]{hyperref}
\usepackage{stfloats}
\usepackage[ruled, vlined, norelsize]{algorithm2e} 
\SetAlFnt{\sf} 

\graphicspath{{incl/}}

\usepackage{soul}
\newcommand{\drop}[1]{\textcolor{red}{#1}}
\renewcommand{\drop}[1]{}

\newdimen\arrayruleHwidth
\makeatletter
\def\Hline{\noalign{\ifnum0=`}\fi\hrule \@height \arrayruleHwidth
\futurelet \@tempa\@xhline}
\makeatother

\makeatletter
\def\blfootnote{\xdef\@thefnmark{}\@footnotetext}
\makeatother

\shortdate
\settimeformat{ampmtime}

\ifCLASSINFOpdf
\else
\fi
\begin{document}
\title{Towards Secure Composition of Integrated Circuits and Electronic Systems: On the Role of EDA}

\author{
Johann Knechtel$^{\textrm 1}$,
Elif Bilge Kavun$^{\textrm 2}$,
Francesco Regazzoni$^{\textrm 3}$,
Annelie Heuser$^{\textrm 4}$,
Anupam Chattopadhyay$^{\textrm 5}$,
\\
Debdeep Mukhopadhyay$^{\textrm 6}$,
Soumyajit Dey$^{\textrm 7}$,
Yunsi Fei$^{\textrm 8}$,
Yaacov Belenky$^{\textrm 9}$,
Itamar Levi$^{\textrm 10}$,
Tim Güneysu$^{\textrm 11}$,
\\
Patrick Schaumont$^{\textrm 12}$,
and
Ilia Polian$^{\textrm 13}$
\\[4pt]
\normalsize
$^{\textrm 1}$johann@nyu.edu -- NYU Abu Dhabi, UAE;
$^{\textrm 2}$e.kavun@sheffield.ac.uk -- University of Sheffield, UK;
\\
$^{\textrm 3}$regazzoni@alari.ch -- ALaRI, University of Lugano, Switzerland;
\\
$^{\textrm 4}$annelie.heuser@irisa.fr -- Univ Rennes, Inria, CNRS, IRISA, France;
\\
$^{\textrm 5}$anupam@ntu.edu.sg -- NTU, Singapore;
$^{\textrm 6, 7}$debdeep@cse.iitkgp.ac.in, soumya@cse.iitkgp.ac.in -- IIT Kharagpur, India;
\\
$^{\textrm 8}$yfei@ece.neu.edu -- NEU, Boston, USA;
$^{\textrm 9}$yaacov.belenky@intel.com -- Intel, Israel;
\\
$^{\textrm 10}$itamar.levi@biu.ac.il -- BIU, Ramat Gan, Israel;
$^{\textrm 11}$tim.gueneysu@rub.de -- RUB \& DFKI, Bochum \& Bremen, Germany;
\\
$^{\textrm 12}$schaum@vt.edu -- VT, Blacksburg, USA;
$^{\textrm 13}$ilia.polian@informatik.uni-stuttgart.de -- University of Stuttgart, Germany
}

\maketitle

\renewcommand{\headrulewidth}{0.0pt}
\thispagestyle{fancy}
\pagestyle{plain}
\cfoot{
	\vspace{-1cm}
\copyright~2020 IEEE.
This is the authors' version of the work. It is posted here for your personal use.
	Not for redistribution.\\
	The definitive Version of Record is published in
	Proc. Design, Automation \& Test in Europe (DATE) 2020\\
}

\begin{abstract}
Modern electronic systems become evermore complex, yet remain modular, with integrated circuits (ICs) acting as versatile hardware components at their heart.
Electronic design automation (EDA) for ICs has focused traditionally on power, performance, and area.
However, given the rise of hardware-centric security threats,
we believe that EDA must also adopt related notions like \textit{secure by design} and \textit{secure composition of hardware}.
Despite various promising studies, we argue that some aspects still require more efforts, for example:
effective means for compilation of assumptions and constraints for security schemes, all the way from the system level down to the ``bare metal'';
modeling, evaluation, and consideration of security-relevant metrics;
or automated and holistic synthesis of various countermeasures, without inducing negative cross-effects.

In this paper, we first introduce hardware security for the EDA community. Next we review prior (academic) art for EDA-driven security evaluation and implementation of countermeasures.
We then discuss
strategies and challenges for advancing research and development toward secure
composition of circuits and systems.

\end{abstract}
\section{Introduction}%
\label{sec:introduction}
Electronic systems are at the heart of our modern societies which are heavily reliant on ubiquitous information technology (IT).
Nowadays, however, an alarmingly large number of security risks are associated with electronic systems.
Ensuring confidentiality, integrity, and availability---the
three key pillars for IT security---directly within the hardware of electronic systems represents a wide-ranging task that is crucial, yet quite challenging.
The related field of hardware security has been driven traditionally by the cryptography community, and rightfully so; the
formal
security promises of any cryptographic algorithm may fail relatively easily once the physical realities of hardware come into play.
For example,
it is well known that cryptographic algorithms leak sensitive information when subjected to side-channel attacks~\cite{brier04} or fault-injection attacks~\cite{journals/pieee/BarenghiBKN12}.
While at least parts of the electronic design automation (EDA) community
have become aware of these and other threats
over the years,
and also proposed some EDA measures to counter them, we argue that more concerted efforts are required.

In this paper, we aim to educate the broader EDA community on the different security threats arising for integrated circuits (ICs) throughout their life cycle, i.e., during design, manufacturing,
and at runtime.
In Table~\ref{tab:threats}, we list the threats covered in this paper and the roles we see for EDA in general.

We motivate in Sec.~\ref{sec:background}, we review the prior art in some detail in Sec.~\ref{sec:prior_art}, and we discuss strategies and challenges for advancements in Sec.~\ref{sec:challenges_strategies}.
Overall, we call for paradigms like \emph{secure by design} and \emph{secure composition of hardware}, i.e., for efforts to account holistically for security notions along with traditional notions of design optimization.

\begin{table}[tb]
\centering
\footnotesize
\setlength{\tabcolsep}{1.6mm}
\caption{Security Threats for ICs and Related Roles of EDA}
\label{tab:threats}

\begin{tabular}{*{3}{c}}

\hline
\textbf{Threat Vector} & 
\textbf{Time of Attack} &
\textbf{Role of EDA} \\
\hline

\multirow{2}{*}{Side-channel attacks}
	& \multirow{2}{*}{Runtime}
	& Evaluation, mitigation
	\\
	&
	& at design time
	\\
\hline

\multirow{2}{*}{Fault-injection attacks}
	& \multirow{2}{*}{Runtime}
	& Evaluation, mitigation
	\\
	&
	& at design time
	\\
\hline

Piracy of design
	& \multirow{2}{*}{Manufacturing;}
	& \multirow{2}{*}{Mitigation}
	\\
intellectual property (IP);
	& \multirow{2}{*}{in the field}
	& \multirow{2}{*}{at design time}
	\\
counterfeiting of ICs
    &
    &
    \\
\hline

\multirow{3}{*}{Hardware Trojans}
	& \multirow{2}{*}{Design;}
	& Mitigation, verification at
	\\
	& \multirow{2}{*}{manufacturing}
	& design time; preparing 
    \\
	& 
	& for testing, inspection 
	\\
\hline

\end{tabular}
\smallerspacecaption
\end{table}

\section{Background and Motivation}
\label{sec:background}
\subsection{Security Threats and Overview on Countermeasures}

Next, we introduce briefly the security aspects we cover in this paper. This section is an overview and not comprehensive; we discuss related prior art in Sec.~\ref{sec:prior_art} in more detail.

\subsubsection{Side-Channel Attacks (SCAs)}
SCAs exploit information leakage from
measurable physical channels and sensitivities of (i) the circuitry itself or (ii) the architecture.
For example, concerning (i), advanced encryption standard (AES) implementations are well-known to be vulnerable to power SCAs  when unprotected \cite{brier04}; concerning (ii), modern microprocessors leak information through timing behaviour of caches, also related to speculative execution \cite{kocher18}.

Most countermeasures apply some kind of ``hiding'' or \emph{masking},
i.e., diffusion of the information leakage, by various means taken across different levels, starting from the system level and ranging down to gates/registers \cite{bellizia18}.
Formal approaches to masking, e.g., \cite{GMK:16}, refer to splitting the computation variables into sections or \emph{shares}
such that internal computations are never performed jointly on all shares.

\subsubsection{Fault-Injection Attacks (FIAs)}
FIAs induce faults to aid in deducing sensitive information.
This includes direct, invasive fault injection, e.g., by laser light \cite{selmke16} or electromagnetic waves \cite{dehbaoui12}, as well as indirect, architectural fault injection, e.g,. by repetitive writing to particular memory locations \cite{vanderVeen16}.

Countermeasures can be separated into detection of FIAs at runtime versus FIA mitigation at design time (e.g.,~\cite{NVKA:19, KGKP:18}).

\subsubsection{Piracy of Design IP, Counterfeiting of ICs}
\label{subsec:piracy_counterfeiting}
Such attacks related to outsourced IC supply chains can be carried out by various adversaries,
ranging from designers, foundry or test facility employees, and even to end-users.

Popular countermeasures against IP piracy are logic locking, split manufacturing, and camouflaging~\cite{knechtel19_IP_COINS}.
Both split manufacturing and camouflaging alter the manufacturing process to protect against untrusted foundries and malicious end-users, respectively.
In contrast, logic locking works at the design level to protect against untrusted foundries and malicious end-users (although the latter relies on tamper-proof memories, which can become targets themselves).
Popular countermeasures against counterfeiting include watermarking and physically-unclonable functions (PUFs)~\cite{rostami14}.

\subsubsection{Hardware Trojans}
Given that IC supply chains are outsourced, adversaries
at various entities could also introduce malicious hardware modifications, known as Trojans.\footnote{Although it has been projected traditionally as the main threat scenario, the likelihood of Trojans being introduced at fabrication time is arguably very low. That is because
any such endeavour, once detected, would fatally disrupt the reputation and business of the related foundry. Therefore, foundries can be expected to employ all organizational and technical means available to hinder unauthorized modifications by malign employees.}
The notion of Trojans is wide-ranging~\cite{BT18}---it describes malicious modifications that are (i) working at the system level, register-transfer level (RTL), gate/transistor level, or the physical level; (ii) seeking to leak information, reduce the IC's performance, or disrupt the IC's working altogether; (iii) are always on, triggered internally, or triggered externally; etc.

Countermeasures can be classified into Trojan detection, conducted pre-silicon and/or post-silicon, and Trojan mitigation. The former relies on testing, verification, and inspection, whereas the latter includes security features
to improve testability/inspection~\cite{BT18} or information-flow tracking~\cite{PWGKR:19}, etc.

\subsection{Classical EDA Flows and Security Fallacies}
\label{subsec:EDA_insecure}

\begin{figure}[tb]
	\centering
	\includegraphics[width=\textwidth]{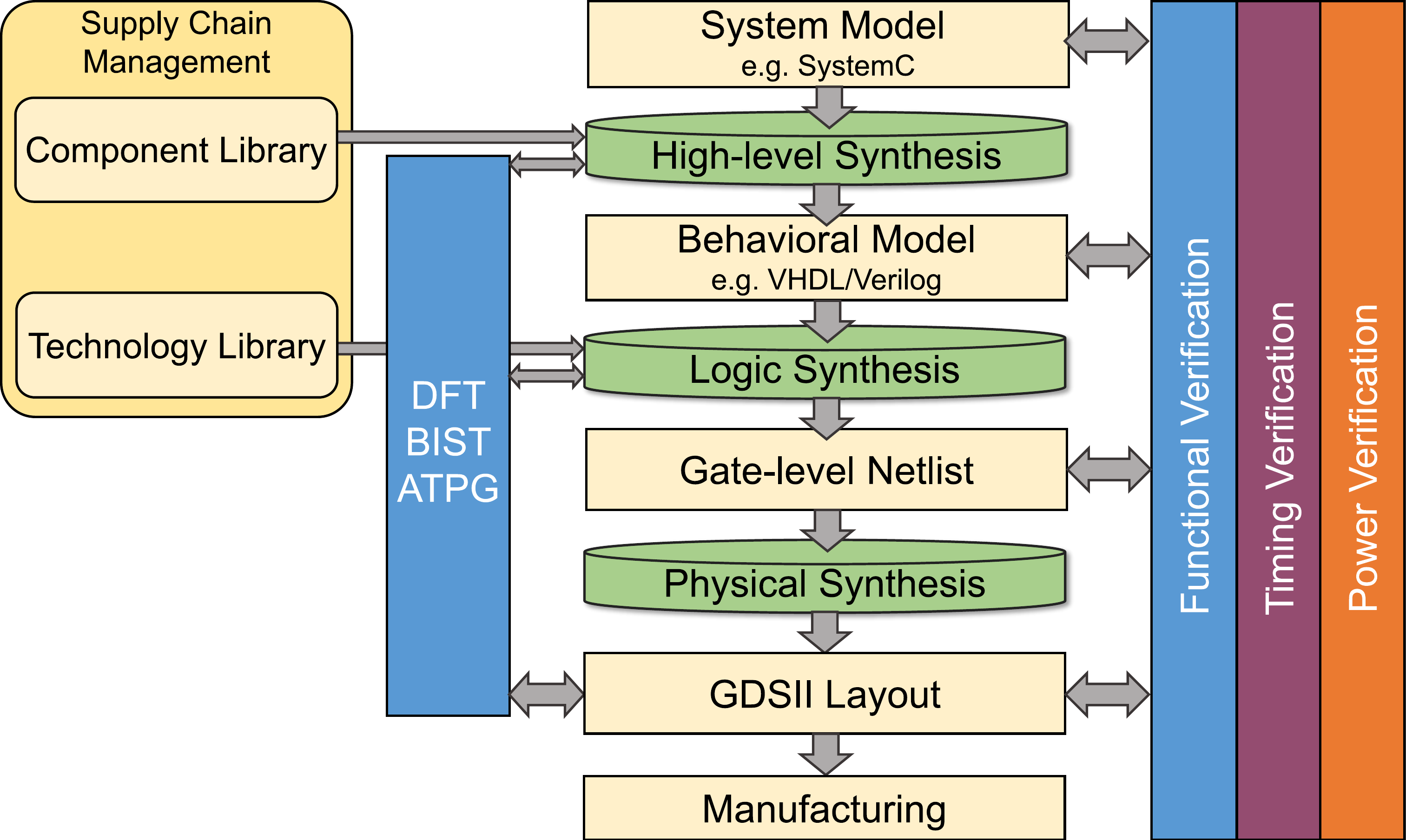}
	\caption{Classical EDA flow, without security considered explicitly.}
	\label{fig:classical_EDA}
\end{figure}

In Fig.~\ref{fig:classical_EDA}, a classical EDA flow is shown in overview.
Various EDA tools as well as design components and technology libraries are involved, which are all provided by potentially malicious third parties.
This presents clearly one of the threats for secure composition of ICs.
For example, Trojans could be introduced directly by adversarial designers, indirectly through untrustworthy third-party IP components, or even through ``hacks'' of EDA tools or the IT environment \cite{BT18}.

State-of-the-art EDA tools provide powerful solutions for simulation, verification, and testing, and they are also well-tailored to optimize any design for power, performance, and area (PPA).
However, these tools are neither tailored yet to account for, e.g., information leakage exploited by SCAs, nor do they offer to incorporate countermeasures in a way that maintains optimization \emph{and} security guarantees.

{\em Motivational example:}
Here we show how classical EDA tools can undermine security. We consider the notion of \emph{private circuits}~\cite{DbRoyICCD} as an example, a scheme that guarantees confidentiality in the face of SCAs in a controllable and quantifiable manner.
Without loss of generality, a bit $a$ of sensitive information can be encoded as a vector $(a_1, a_2, a_3)$, where $a=a_1 \oplus a_2 \oplus a_3$ and
$\oplus$ denotes bitwise XOR.
Any regular operations are implemented in encoded form and incorporate random bits $r_{i,j}$, where $1 \leq i, j \leq 3$.
For example, the AND operation $c = a \wedge b$ on such vectors is computed as: $c_1=a_1b_1 \oplus r_{1,2}\oplus r_{1,3}$ and $c_2=a_2b_2 \oplus (r_{1,2}\oplus a_1b_2)\oplus a_2b_1 \oplus r_{2,3}$  and $c_3=a_3b_3\oplus (r_{1,3}\oplus a_1b_3)\oplus a_3b_1 \oplus (r_{2,3} \oplus a_2b_3) \oplus a_3b_2$.

\begin{figure}[tb]  
	\centering
	\includegraphics[width=.85\textwidth]{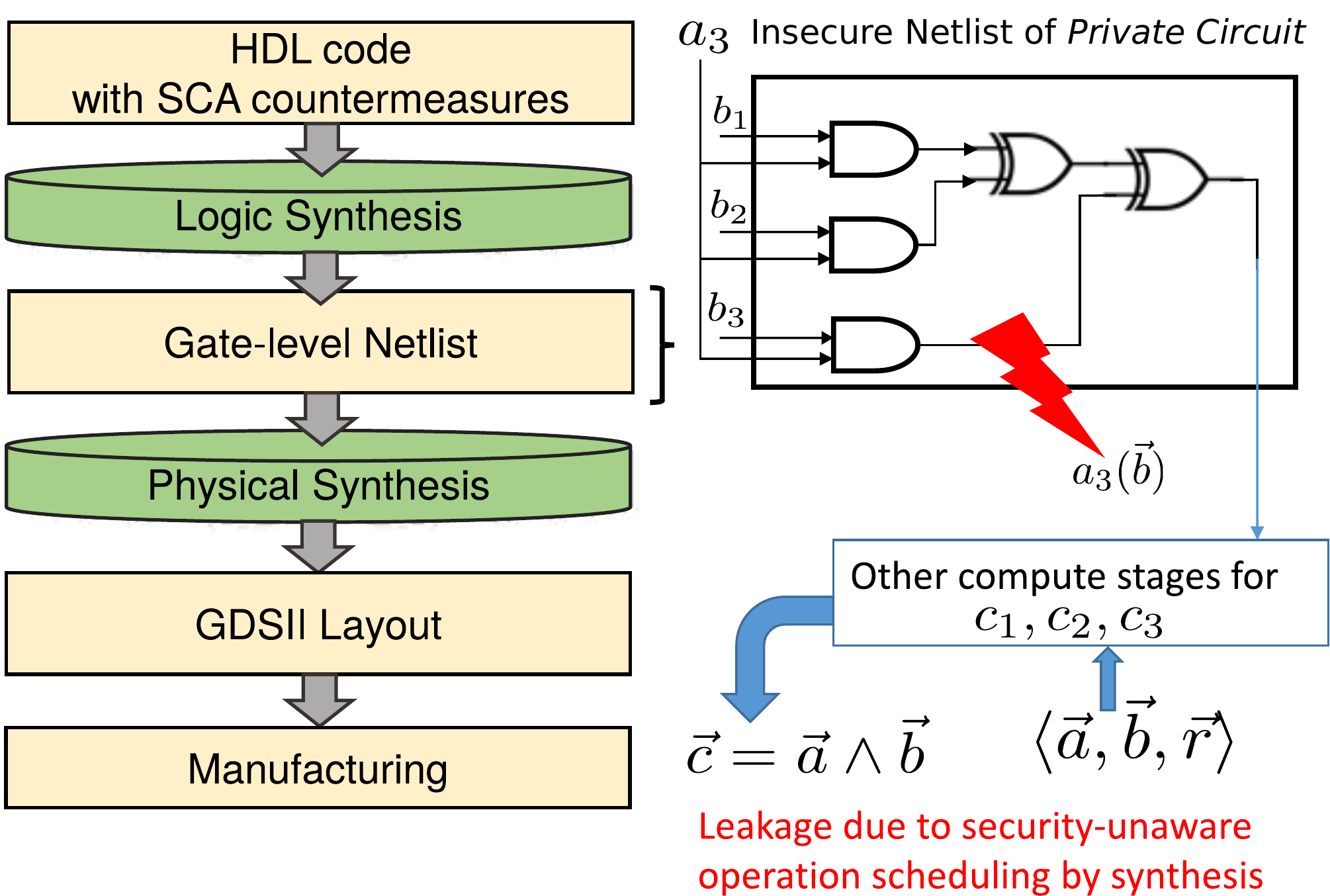}
	\caption{Motivational example for the insecure nature of classical EDA tools.}
	\label{fig:insecure_EDA__private_circuit}
\end{figure}

The security promise by private circuits is based on the fact that all components of one such vector are never processed at the same time. Thus, an adversary cannot learn it from power measurements (or other side channels).
Now, it is important to note that
the order of computation, as indicated by parentheses, is critical for suppressing information leakage, even though it is irrelevant for correctness (as $\oplus$ is commutative).
For the example of the AND operation, let us assume the synthesis tool implements $c_3$ such that the expression $a_3b_1\oplus a_3b_2 \oplus a_3b_3=a_3(b)$ is derived first and the random bits $r_{ij}$
are added only later, then the computation will leak the value of $b$
(Fig.~\ref{fig:insecure_EDA__private_circuit}).
Regular, security-unaware tools may take such decisions easily, e.g., when it helps to improve timing.

Note that leakage can occur even when private circuits are synthesized in a security-aware manner, e.g., then due to delays and glitches for the random variables. An effective and well-known, yet limited, approach for leakage evaluation is \emph{test vector leakage assessment (TVLA)}~\cite{tvla}; see Sec.~\ref{sec:prior_art}.

\subsection{Challenges and Tasks for Security-Centric EDA}

As listed in Table~\ref{tab:threats}, we see potential for EDA tools to evaluate and mitigate various threats already at design time.
Considering though the vastly different nature of these threats,
it may seem impossible to provide
comprehensive security-centric EDA flows. However, to make progress towards this ultimate goal,
we argue that the EDA community should (continue to) focus on key 
challenges which are, among others:
\begin{itemize}
    \item Evaluation and consideration of security-relevant metrics, with varying level of detail for different EDA stages;
    \item Effective means for compilation of assumptions and constraints for security schemes, all the way from the system level down to the ``bare metal'';
    \item Automated and holistic synthesis of various countermeasures, without inducing negative cross-effects.
\end{itemize}
We believe that the community is actually
well-positioned to address these challenges. EDA tools are driven by metrics and heuristics and also have to tackle trade-offs continuously
while step-wise refining the design quality---these
principles can certainly be extended towards secure composition of ICs~\cite{RAVI201917}.

For any security scheme, it is essential to first define the \emph{threat model}, which describes the adversary's assets, capabilities, constraints, and goals,
along with the proposed countermeasures.
Depending on the type and time of attack, doing so can become quite complex; e.g., for SCAs and FIAs at runtime, many physical aspects come into play, like means of fault injection, temperature, voltage,
power-supply impedance, glitches, etc.
Even once threat models are defined properly---and translated into specific metrics and countermeasures which can be handled by EDA tools---they can still have significant weak spots. For one, it is impossible to hinder an adversary from taking further efforts going beyond the modeled means of attack.
For another, incorporating the threat model into the EDA tools
will be subject to inaccuracies, not only due to limitations
on computational cost when exploring the design space of complex ICs with all regular components and various security features, but also due to limitations of metrics and evaluation techniques themselves.

It is understood that EDA tools cannot provide perfect security but it is an essential task to formulate and explore the practical bounds for security schemes when embedded in hardware.
EDA tools should assist the designer with automated integration of security features and countermeasures but also needs to formulate the related limitations and remaining risks clearly, to enable effective risk management. 

\section{Discussion of Prior Art}
\label{sec:prior_art}
\begin{table*}[tb]
\centering
\footnotesize
\caption{Security Schemes Suitable for Incorporation into EDA Tools}
\label{tab:edaprot}

\begin{tabular}{@{}ccccc@{}}

\hline
\multirow{2}{*}{\bf Design Stage}
& \multicolumn{4}{c}{\bf Threat Vectors} \\
\cline{2-5}
& \bf Side-Channel Attacks & \bf Fault-Injection Attacks
& \bf IP Piracy and Counterfeiting & \bf Trojans \\
\hline
& Information-flow tracking \cite{PWGKR:19};
& \multirow{2}{*}{Error-detecting architectures \cite{KGKP:18};} &  \multirow{2}{*}{Metering IP} &  \\
High-level synthesis & Integration of masking \cite{GMK:16}; & \multirow{2}{*}{Infective countermeasures \cite{PCM:17}} & \multirow{2}{*}{(including PUFs) \cite{AK:07}} & Self-authentication \cite{XT:13} \\
& Register flushing & \\
\hline
\multirow{2}{*}{Logic synthesis}	& Gate-level
protections \cite{TV:06};
& \multirow{2}{*}{Automatic fault analysis \cite{BHB:19}} & Camouflaging \cite{RPSK:12}; & Automatic insertion of \\
& Identification of leaking gates & & Logic locking \cite{RKM+:08} & security monitors \cite{WGH+:16} \\
\hline
Physical synthesis & Low-level information leakage & Embedding sensors \cite{NVKA:19,DFARPA};
& Split manufacturing \cite{mccants16}; & \multirow{2}{*}{Embedding sensors \cite{ZFT:13}} \\
(place and route) & analysis (TVLA \cite{tvla}, etc.) & Shielding \cite{LM:06} & Entropy primitives \cite{GDT:18} &\\
\hline
Functional & Identification of architectural & Validation of error-detection
& Correctness of locked logic; & Proof-carrying
\\
validation & covert channels \cite{FSBMK:19} & properties \cite{FSFD:11} & De-obfuscation attacks \cite{AKHS:19} & hardware \cite{LJM:12}
\\
\hline
Timing and power & Pre-silicon power/timing & Detailed modeling of & Validation of low-level & \multirow{2}{*}{Fingerprinting \cite{JM:08}} \\
verification & simulation \cite{JAC+:13,he19} & fault injections \cite{DBC+:18} & properties of PUFs & \\
\hline
Testing (ATPG, & Securing DFT against read-out & DFX architecture to handle & IP protection integrated & Pattern generation for  \\
DFT, BIST) & (scan-chain attacks \cite{YWK:05}, etc.) & malicious/natural failures & into DFX infrastructure & Trojan detection \cite{CWPPB:09} \\

\hline
\end{tabular}
\smallerspacecaption
\end{table*}

Table \ref{tab:edaprot} summarizes security schemes that could be (and partially are already)
supported by EDA tools, categorized into design stages versus threat vectors.\footnote{There
is a significant number of studies for most
of the table's entries, but we can focus only on selected works within the page limits.}
Some schemes are based on a ``red team versus blue team'' approach, i.e., they leverage the relevant attack(s) internally,
with the objective to
inform the designer how to address remaining vulnerabilities or, more challenging, to demonstrate the absence of vulnerabilities.
For example, to demonstrate whether an error-detecting
scheme can detect all faults
means to search for other (types of) faults that are possibly missed.
A different approach is to quantify threats through evaluation of metrics, but without considering an explicit attack scenario. For example, countermeasures against SCAs are often assessed by
information leakage via statistical or information-theoretical procedures.

In the following, we provide a brief overview and discussion of prior art for each row of Table \ref{tab:edaprot}.

\subsection{Security-Driven High-Level Synthesis}

High-level synthesis (HLS) allocates IP blocks and functional units, binds tasks to these components, and schedules the task execution.
An HLS tool would ideally allocate IP blocks 
automatically as needed for various security-related tasks: (i) secure random number generators (RNGs) \cite{Sch:09}
for key generation or masking, (ii) PUFs
for circuit identification, authentication, and metering \cite{RH:14,AK:07},
(iii) self-authentication
logic \cite{XT:13} or wrapper architectures \cite{basak17}
to complicate insertion of Trojans,
(iv) error-detecting or shielding architectures against
FIAs \cite{KGKP:18,PCM:17}, etc.
Another simple countermeasure against SCAs
could be to instruct HLS to randomly flush/overwrite registers holding critical data after their use.

While there are works on automated synthesis of masking for software \cite{EW:14}, EDA-centric approaches for hardware are still in development.
Towards this end, formalized security requirements are an important input for security-centric HLS tools. These can be specified in
secure hardware languages like Caisson \cite{cai} or SecVerilog \cite{sec}. Another language called QIF-Verilog \cite{GDHTJ:19} provides the techniques of quantitative information flow (QIF) tracking \cite{MAHC:14} in the hardware domain. In general, techniques for 
information-flow tracking developed in the context of software engineering \cite{vmcai} can also be used to validate the 
resilience of RTL code resulting from
HLS tools \cite{PWGKR:19}. The method reported in \cite{vmcai} leverages  approximate model counting in order to handle large program state spaces, a concept which is also useful in the context of information-flow tracking for practical EDA  use cases.

\subsection{Security-Driven Logic Synthesis}

Logic synthesis is the step of compiling the high-level RTL into an actual netlist, mapping it
to the technology of choice. This can be combined with gate-level security schemes, e.g., to reduce information leakage exploited by SCAs following the wave dynamic differential logic (WDDL) paradigm \cite{TV:06}.
Such ``hiding'' schemes represent 
alternatives or complements to formal masking approaches.
Methods for automatic fault analysis, some also suitable for logic synthesis, are reviewed in \cite{BHB:19}.
Moreover, logic synthesis can be tasked to instantiate security monitors to help detecting Trojans at runtime \cite{WGH+:16}.

Concerning IP protection, two approaches are applicable here,
camouflaging and logic locking. 
Logic synthesis has to employ camouflaging according to the scale desired by the
designer, where synthesis is constrained to the Boolean functionalities covered by the multi-functional but obfuscated primitives---this is similar to regular but constrained synthesis and is well supported.
For locking, however, there is a need to support security requirements formulated at the behavioral level. Currently, locking is implemented directly at the gate-level netlist.
Similar to the example for private circuits in Sec.~\ref{subsec:EDA_insecure}, synthesis is unaware of the security notion for locking. Thus, among others, locking is prone to structural attacks targeting at the
synthesized (or layout-level) netlist~\cite{chakraborty18_SAIL,yang19}.

\subsection{Security-Driven Physical Synthesis}

Physical synthesis is the step of generating an optimized design from the gate-level netlist, through means of place and route (PnR), clock-tree design, timing closure, etc.

Forming the key step towards the ``bare metal,'' it is crucial that physical synthesis considers security notions that are primarily subject to physical phenomena. For example, an important task here is to 
quantify the information that is leaked through the various side-channels of an IC.
The most relevant approach for such evaluation is TVLA~\cite{tvla}. In general, TVLA uses \emph{Welch's t-test statistics} to quantify the differences between the means of two data sets describing some physical phenomena.
The validity of TVLA for evaluating SCA resilience is subject to the assumptions made in the threat model,
like the noise distribution assumed for the measurements taken by the attacker.
Information-theoretic procedures can bound that error using fewer statistical assumptions, but they require careful characterization of
the side-channel probability distribution, which is computationally costly
(since Maxwell's equations are to be tackled).
In any case, most of these metrics and procedures are challenged by the fact that information leakage is multi-dimensional and multi-variate.

Physical design could also be tailored to employ security primitives like RNGs \cite{Sch:09} or PUFs \cite{RH:14,AK:07}, shields to protect against FIAs \cite{LM:06}, sensors
to detect FIAs \cite{NVKA:19,DFARPA}, or Trojan detection circuitry \cite{ZFT:13}. Since the entropy harnessed by on-chip RNGs and PUFs comes from physical circuit structures, layout-level optimization of their properties is required \cite{GDT:18}.

As for IP protection, split manufacturing is to be supported at this stage.
The security promise of split manufacturing---foremost to hinder IP piracy, but also Trojan insertion, both conducted by foundry adversaries---relies on providing a ``meaningless sea of gates with dangling wires'' to the untrusted foundry (the higher metal layers are manufactured subsequently by another, trusted facility).
Classical EDA flows work holistically on the IC stack, leaving layout-level hints for adversaries, e.g., equipped with machine learning~\cite{li19_SM_ML_DAC}.
Thus, it is essential for split manufacturing that physical synthesis is tailored to dissolve such hints (yet optimize for PPA). This can be achieved, e.g., by selective ``pushing'' of wires to the higher metal layers~\cite{patnaik18_SM_ASPDAC} or by placement perturbation~\cite{SPK+:17}.

\subsection{Security-Driven Functional Validation}

Validation covers simulation and formal techniques,
including equivalence and property checking.
Especially the latter is helpful for analysis of security properties and proving their effectiveness.
For one, a recent study uses formal methods to identify architectural vulnerabilities in advanced microprocessors \cite{FSBMK:19}.
For another, when verifying
an error-detection architecture, i.e., when checking for fault coverage,
formal analysis developed
for transient faults can play a role \cite{FSFD:11}.

For IP protection, verification serves to check the correctness of logic modifications introduced by locking or camouflaging. More importantly even, verification can be used to mimic attackers leveraging satisfiability-based tools (i.e., SAT and SMT solvers), and to demonstrate the resilience of protected ICs against such powerful attacks \cite{AKHS:19}.

Finally, concerning Trojans, 
security properties can be embedded directly in the HDL/RTL to obtain ``proof-carrying hardware'' \cite{LJM:12}.
Such schemes should be integrated into property-checker flows and tools.

\subsection{Security-Driven Timing and Power Verification}

Timing and power verification serves to achieve design closure.
One distinguishes between
simulation of timing/power artifacts and ``vectorless'' analytical approaches, e.g., proving
that IR-drop
will not exceed a given limit.
Simulation approaches are particularly useful for analysis of information leakage through side channels;
it is desirable to identify such leakage (or demonstrate its absence) through pre-silicon simulations rather
than belatedly measure the final, manufactured ICs.
Pre-silicon simulations may also point to the origin of information leakage in the circuitry, thus enabling the designer to fix the underlying problem.

Existing simulation tools work on different abstraction levels and support different degrees of accuracy,
from detailed SPICE analysis to fast gate-level approaches.
A critical detail for simulation is timing and gate delays;
it has been reported that
glitches (i.e., transient signals within a clock cycle) influence information leakage \cite{MMSS:19}.
Whether glitches remain present in the actual IC, however, depends on physical-synthesis results, manufacturing variability, and also ambient conditions.
It seems an open question how accurate the timing/power models used for simulation must be to obtain reliable prediction
about the expected information leakage.

Timing/power verification is also the stage
to run detailed analysis of fault injections using accurate electrical models \cite{DBC+:18}, or to verify the
behavior of PUFs in terms of entropy, reliability, and uniqueness.
Simulation also serves well for \emph{fingerprinting} \cite{JM:08}, a countermeasure against Trojans, which is based on checking consistency of path delays.

\subsection{Security-Driven Testing}

Testability is crucial, yet contradictory to security to some degree \cite{Pol:14}.
That is because
test, diagnosis, and debug features enable comprehensive access to IC internals, providing
an attacker the opportunity to read out sensitive information (e.g., via scan-based attacks \cite{YWK:05}).
As a consequence, emerging test standards will also incorporate security measures \cite{VSNFR:19}.

More complex ICs incorporate a ``design for X'' (DFX) infrastructure, which combines classical scan-based testing with build-in self test (BIST) features for logic and memory, transient-fault detection and re-configuration logic,
circuitry for yield management, and debug and diagnostic features \cite{PAA+:08}.
To integrate FIA detection into the same DFX infrastructure seems only logical.
However, distinguishing between natural and malicious faults is non-trivial \cite{KGKP:18b}, and the responses should be different: fastest possible recovery and resumption
of regular operation upon a natural fault, but re-keying or even discontinuation of service upon a tampering attempt.
Therefore, future security-aware DFX infrastructures should enable such distinction. Besides, they may also manage IP protection techniques, e.g., for key management for locking.

There is extensive prior art for detecting Trojans through means of testing.
This covers (i) functional tests that aim at triggering Trojans \cite{CWPPB:09} and (ii) parametric tests that aim at detecting Trojans' fingerprints through side-channels \cite{AARP:10}.
While such tests can be included into automatic test pattern generation (ATPG) tools, their
effectiveness in reliably identifying strategically hidden Trojans in large and complex
ICs remains to be proven.

\section{%
Strategies and Challenges Towards Secure Composition Using EDA Tools}
\label{sec:challenges_strategies}
Security of any system is subject to its weakest link, and ICs form no exception here.
We have covered, in overview,
the large variety of hardware-related threats and 
countermeasures, along with some discussion of current limitations.

It is known that not all types or implementations of countermeasures
are \emph{composable}, e.g.,
adding error-detecting logic can deteriorate resilience against SCAs \cite{RBIK:12}.
Thus, tools for joint compilation of countermeasures and, even more importantly, for verifying their effectiveness are required.
Ideally, once the security-enforcing designers have implemented yet another countermeasure, they can re-run the envisioned security-centric EDA flow which then covers all threats, also seemingly unrelated ones, to hinder that any countermeasure has become inadvertently compromised.

To become a reality, such security-centric EDA tools require effective and efficient security metrics and evaluation techniques.
The whole EDA domain is metrics-driven, and EDA tools are well positioned to balance between, e.g., a circuit's
area and testability, all
quantified by meaningful metrics.

While several security metrics and evaluation techniques are known \cite{rostami14}, as also outlined in this paper,
the necessary assumption of an intelligent and strategic attacker complicates their definition and usage.
For example, a transient fault that leads to a critical system failure can be ignored during
reliability analysis in case it is extremely unlikely to occur. When it comes to resistance against
FIAs, however, the attacker may put extra effort into injecting precisely this fault; it
cannot be ignored anymore.
Having to account for such ``unlikely but possible'' events
poses a significant burden for security analysis and on appropriate strategies to incorporate such analysis into EDA tools.
This also implies that
one can expect some security metrics to act more like step functions,
where certain efforts must be spent to reach a security level,
but spending more will not provide additional benefits. This is
fundamentally different from classical metrics like area or power consumption and
should be considered accordingly for security-aware design space exploration.

\section{Conclusion}
\label{sec:conclusion}
EDA tools are traditionally a key enabler for complex ICs and electronic systems.
Nowadays, an increasing number of applications becomes security-critical and ICs must offer protection against hardware-oriented attacks, yet the support by EDA tools is lacking for this matter.
We outlined short- to medium-term potentials
for security-driven design methods
to be integrated into EDA tools. We also identified
conceptual challenges for secure composition of countermeasures against various threat vectors and for security metrics.

\section*{Acknowledgements}
This work originates from
Dagstuhl Seminar 19301,
\textit{Secure Composition for Hardware Systems},
July 21--26, 2019.

\bibliographystyle{IEEEtran}
\bibliography{DATE}

\begin{thebibliography}{10}
\providecommand{\url}[1]{#1}
\csname url@samestyle\endcsname
\providecommand{\newblock}{\relax}
\providecommand{\bibinfo}[2]{#2}
\providecommand{\BIBentrySTDinterwordspacing}{\spaceskip=0pt\relax}
\providecommand{\BIBentryALTinterwordstretchfactor}{4}
\providecommand{\BIBentryALTinterwordspacing}{\spaceskip=\fontdimen2\font plus
\BIBentryALTinterwordstretchfactor\fontdimen3\font minus
  \fontdimen4\font\relax}
\providecommand{\BIBforeignlanguage}[2]{{%
\expandafter\ifx\csname l@#1\endcsname\relax
\typeout{** WARNING: IEEEtran.bst: No hyphenation pattern has been}%
\typeout{** loaded for the language `#1'. Using the pattern for}%
\typeout{** the default language instead.}%
\else
\language=\csname l@#1\endcsname
\fi
#2}}
\providecommand{\BIBdecl}{\relax}
\BIBdecl

\bibitem{brier04}
E.~Brier \emph{et~al.}, ``{Correlation Power Analysis with a Leakage Model},''
  \emph{{CHES}}, vol. 3156, pp. 16--29, 2004.

\bibitem{journals/pieee/BarenghiBKN12}
A.~Barenghi \emph{et~al.}, ``{Fault Injection Attacks on Cryptographic Devices:
  Theory, Practice, and Countermeasures},'' \emph{Proceedings of the IEEE},
  vol. 100, no.~11, pp. 3056--3076, 2012.

\bibitem{kocher18}
P.~Kocher \emph{et~al.}, ``{Spectre Attacks: Exploiting Speculative
  Execution},'' \emph{SP}, vol.~1, pp. 19--37, 2019.

\bibitem{bellizia18}
D.~Bellizia \emph{et~al.}, ``{Secure Double Rate Registers as an RTL
  Countermeasure Against Power Analysis Attacks},'' \emph{{TVLSI}}, vol. 26-7,
  2018.

\bibitem{GMK:16}
H.~Gro{\ss} \emph{et~al.}, ``{Domain-Oriented Masking: Compact Masked Hardware
  Impl. with Arbitrary Protection Order},'' in \emph{TIS@CCS}.\hskip 1em plus
  0.5em minus 0.4em\relax {ACM}, 2016.

\bibitem{selmke16}
B.~Selmke \emph{et~al.}, ``{Attack on a DFA Protected AES by Simultaneous Laser
  Fault Injections},'' in \emph{{FDTC}}, 2016, pp. 36--46.

\bibitem{dehbaoui12}
A.~Dehbaoui \emph{et~al.}, ``{Injection of transient faults using
  electromagnetic pulses Practical results on a cryptographic system},'' in
  \emph{{ePrint-123}}, 2012.

\bibitem{vanderVeen16}
V.~van~der Veen \emph{et~al.}, ``{Drammer: Deterministic Rowhammer Attacks on
  Mobile Platforms},'' in \emph{{CCS}}, 2016, pp. 1675--1689.

\bibitem{NVKA:19}
G.~D. Natale \emph{et~al.}, ``{Hidden-Delay-Fault Sensor for Test, Reliability
  and Security},'' in \emph{{DATE}}.\hskip 1em plus 0.5em minus 0.4em\relax
  {IEEE}, 2019, pp. 316--319.

\bibitem{KGKP:18}
B.~Karp \emph{et~al.}, ``{Security-oriented Code-based Architectures for
  Mitigating Fault Attacks},'' in \emph{{DCIS}}.\hskip 1em plus 0.5em minus
  0.4em\relax {IEEE}, 2018, pp. 1--6.

\bibitem{knechtel19_IP_COINS}
J.~Knechtel \emph{et~al.}, ``Protect your chip design intellectual property: An
  overview,'' in \emph{{COINS}}, 2019, pp. 211--216.

\bibitem{rostami14}
M.~Rostami \emph{et~al.}, ``{A Primer on Hardware Security: Models, Methods,
  and Metrics},'' \emph{{JProc}}, vol. 102, no.~8, pp. 1283--1295, 2014.

\bibitem{BT18}
S.~Bhunia \emph{et~al.}, Eds., \emph{{The Hardware Trojan War: Attacks, Myths,
  and Defenses}}.\hskip 1em plus 0.5em minus 0.4em\relax Springer, 2018.

\bibitem{PWGKR:19}
C.~Pilato \emph{et~al.}, ``{TaintHLS: High-Level Synthesis for Dynamic
  Information Flow Tracking},'' \emph{{IEEE} {TCAD}}, vol. 38-5, p. 798, 2019.

\bibitem{DbRoyICCD}
D.~B. Roy \emph{et~al.}, ``{From Theory to Practice of Private Circuit: A
  Cautionary Note},'' in \emph{ICCD}, 2015, pp. 296--303.

\bibitem{tvla}
J.~Cooper \emph{et~al.}, ``{Test Vector Leakage Assessment (TVLA) Methodology
  in Practice},'' in \emph{International Cryptographic Module Conference},
  2013.

\bibitem{RAVI201917}
P.~Ravi \emph{et~al.}, ``Security is an architectural design constraint,''
  \emph{MICPRO}, vol.~68, pp. 17 -- 27, 2019.

\bibitem{PCM:17}
S.~Patranabis \emph{et~al.}, ``{Fault Tolerant Infective Countermeasure for
  AES},'' \emph{J. Hardware and Systems Security}, vol.~1, no.~1, pp. 3--17,
  2017.

\bibitem{AK:07}
Y.~Alkabani \emph{et~al.}, ``{Active Hardware Metering for Intellectual
  Property Protection and Security},'' in \emph{{USENIX} Security Symposium},
  2007.

\bibitem{XT:13}
K.~Xiao \emph{et~al.}, ``{BISA: Built-in Self-authentication for Preventing
  Hardware Trojan Insertion},'' in \emph{{HOST}}.\hskip 1em plus 0.5em minus
  0.4em\relax {IEEE}, 2013, pp. 45--50.

\bibitem{TV:06}
K.~Tiri \emph{et~al.}, ``{A Digital Design Flow for Secure Integrated
  Circuits},'' \emph{{IEEE} T.{CAD} of ICs and Systems}, vol.~25, no.~7, pp.
  1197--1208, 2006.

\bibitem{BHB:19}
J.~Breier \emph{et~al.}, Eds., \emph{{Automated Methods in Cryptographic Fault
  Analysis}}.\hskip 1em plus 0.5em minus 0.4em\relax Springer, 2019.

\bibitem{RPSK:12}
J.~Rajendran \emph{et~al.}, ``{Security Analysis of Logic Obfuscation},'' in
  \emph{{DAC}}.\hskip 1em plus 0.5em minus 0.4em\relax {ACM}, 2012, pp. 83--89.

\bibitem{RKM+:08}
J.~A. Roy \emph{et~al.}, ``{EPIC: Ending Piracy of Integrated Circuits},'' in
  \emph{{DATE}}.\hskip 1em plus 0.5em minus 0.4em\relax {ACM}, 2008, pp.
  1069--1074.

\bibitem{WGH+:16}
T.~F. Wu \emph{et~al.}, ``{TPAD: Hardware Trojan Prevention and Detection for
  Trusted Integrated Circuits},'' \emph{{TCAD}}, vol.~35, no.~4, pp. 521--534,
  2016.

\bibitem{DFARPA}
M.~{Khairallah} \emph{et~al.}, ``{DFARPA}: Differential fault attack resistant
  physical design automation,'' in \emph{{DATE}}, 2018, pp. 1171--1174.

\bibitem{mccants16}
\BIBentryALTinterwordspacing
C.~McCants, ``Trusted integrated chips ({TIC}) program,'' IARPA, Tech. Rep.,
  2016. [Online]. Available:
  \url{https://www.ndia.org/-/media/sites/ndia/meetings-and-events/divisions/systems-engineering/past-events/trusted-micro/2016-august/mccants-carl.ashx}
\BIBentrySTDinterwordspacing

\bibitem{ZFT:13}
X.~Zhang \emph{et~al.}, ``{Detection of Trojans Using A Combined Ring
  Oscillator Network and Off-chip Transient Power Analysis},'' \emph{{JETC}},
  vol.~9, no.~3, pp. 25:1--25:20, 2013.

\bibitem{LM:06}
H.~Li \emph{et~al.}, ``{Security Evaluation At Design Time Against Optical
  Fault Injection Attacks},'' \emph{IEE Proc.-Inf. Security}, vol. 153-1, pp.
  3--11, 2006.

\bibitem{GDT:18}
Y.~Guo \emph{et~al.}, ``{Variation Enhancement of Arbiter PUFs with Asymmetric
  Layout},'' in \emph{{MWSCAS}}.\hskip 1em plus 0.5em minus 0.4em\relax {IEEE},
  2018, pp. 841--844.

\bibitem{FSBMK:19}
M.~R. Fadiheh \emph{et~al.}, ``{Processor Hardware Security Vulnerabilities and
  their Detection by Unique Program Execution Checking},'' in
  \emph{{DATE}}.\hskip 1em plus 0.5em minus 0.4em\relax {IEEE}, 2019, pp.
  994--999.

\bibitem{FSFD:11}
G.~Fey \emph{et~al.}, ``{Effective Robustness Analysis Using Bounded Model
  Checking Techniques},'' \emph{{TCAD}}, vol.~30, no.~8, pp. 1239--1252, 2011.

\bibitem{AKHS:19}
K.~Z. Azar \emph{et~al.}, ``{SMT Attack: Next Generation Attack on Obfuscated
  Circuits with Capabilities and Performance Beyond the SAT Attacks},''
  \emph{{IACR} TCHES}, vol. 2019, no.~1, pp. 97--122, 2019.

\bibitem{LJM:12}
E.~Love \emph{et~al.}, ``{Proof-Carrying Hardware Intellectual Property: {A}
  Pathway to Trusted Module Acquisition},'' \emph{{IEEE} Trans. Information
  Forensics and Security}, vol.~7, no.~1, pp. 25--40, 2012.

\bibitem{JM:08}
Y.~Jin \emph{et~al.}, ``{Hardware Trojan Detection Using Path Delay
  Fingerprint},'' in \emph{{HOST}}.\hskip 1em plus 0.5em minus 0.4em\relax
  {IEEE} Computer Society, 2008, pp. 51--57.

\bibitem{JAC+:13}
J.~Jiang \emph{et~al.}, ``{MIRID: Mixed-Mode IR-Drop Induced Delay
  Simulator},'' in \emph{Asian Test Symposium}.\hskip 1em plus 0.5em minus
  0.4em\relax {IEEE} ComSoc, 2013, pp. 177--182.

\bibitem{he19}
M.~T. He \emph{et~al.}, ``{RTL-PSC}: Automated power side-channel leakage
  assessment at register-transfer level,'' \emph{arXiv}, 2019.

\bibitem{DBC+:18}
J.~Dutertre \emph{et~al.}, ``{Laser Fault Injection at the CMOS 28 nm
  Technology Node: an Analysis of the Fault Model},'' in \emph{{FDTC}}.\hskip
  1em plus 0.5em minus 0.4em\relax {IEEE} ComSoc, 2018.

\bibitem{YWK:05}
B.~Yang \emph{et~al.}, ``{Secure Scan: A Design-for-test Architecture for
  Crypto Chips},'' in \emph{{DAC}}.\hskip 1em plus 0.5em minus 0.4em\relax
  {ACM}, 2005, pp. 135--140.

\bibitem{CWPPB:09}
R.~S. Chakraborty \emph{et~al.}, ``{MERO: A Statistical Approach for Hardware
  Trojan Detection},'' in \emph{{CHES}}, vol. 5747.\hskip 1em plus 0.5em minus
  0.4em\relax Springer, 2009, pp. 396--410.

\bibitem{Sch:09}
W.~Schindler, ``{Random Number Generators for Cryptographic Applications},'' in
  \emph{Cryptographic Engineering}.\hskip 1em plus 0.5em minus 0.4em\relax
  Springer, 2009, pp. 5--23.

\bibitem{RH:14}
U.~R{\"{u}}hrmair \emph{et~al.}, ``{PUFs At A Glance},'' in \emph{{DATE}},
  2014, pp. 1--6.

\bibitem{basak17}
A.~Basak \emph{et~al.}, ``{Security Assurance for System-on-Chip Designs With
  Untrusted IPs},'' \emph{IEEE TIFS}, vol.~12, no.~7, pp. 1515--1528, 2017.

\bibitem{EW:14}
H.~Eldib \emph{et~al.}, ``{Synthesis of Masking Countermeasures Against Side
  Channel Attacks},'' in \emph{{CAV}}, vol. 8559.\hskip 1em plus 0.5em minus
  0.4em\relax Springer, 2014, pp. 114--130.

\bibitem{cai}
X.~Li \emph{et~al.}, ``{Caisson: A Hardware Description Language for Secure
  Information Flow},'' in \emph{ACM SIGPLAN PLDI}, 2011, pp. 109--120.

\bibitem{sec}
D.~Zhang \emph{et~al.}, ``{A Hardware Design Language for Timing-Sensitive
  Information-Flow Security},'' \emph{SIGPLAN}, vol. 50-4, pp. 503--516, 2015.

\bibitem{GDHTJ:19}
X.~Guo \emph{et~al.}, ``{QIF-Verilog: Quantitative Information-Flow based
  Hardware Description Languages for Pre-Silicon Security Assessment},'' in
  \emph{{HOST}}.\hskip 1em plus 0.5em minus 0.4em\relax {IEEE}, 2019, pp.
  91--100.

\bibitem{MAHC:14}
P.~Mardziel \emph{et~al.}, ``{Quantifying Information Flow for Dynamic
  Secrets},'' in \emph{{IEEE} SP}, 2014, pp. 540--555.

\bibitem{vmcai}
F.~Biondi \emph{et~al.}, ``{Scalable Approximation of Quantitative Information
  Flow in Programs},'' in \emph{Proc. of {VMCAI} Conference}, 2018, pp. 71--93.

\bibitem{chakraborty18_SAIL}
P.~Chakraborty \emph{et~al.}, ``{SAIL: Machine Learning Guided Structural
  Analysis Attack on Hardware Obfuscation},'' in \emph{{AHOST}}, 2018, pp.
  56--61.

\bibitem{yang19}
F.~Yang \emph{et~al.}, ``{Stripped Functionality Logic Locking with Hamming
  Distance Based Restore Unit ({SFLL-hd}) -- Unlocked},'' \emph{{TIFS}}, 2019.

\bibitem{li19_SM_ML_DAC}
H.~Li \emph{et~al.}, ``{Attacking Split Manufacturing from a Deep Learning
  Perspective},'' in \emph{{DAC}}, 2019, pp. 135:1--135:6.

\bibitem{patnaik18_SM_ASPDAC}
S.~Patnaik \emph{et~al.}, ``Concerted wire lifting: Enabling secure and
  cost-effective split manufacturing,'' in \emph{ASPDAC}, 2018, pp. 251--258.

\bibitem{SPK+:17}
A.~Sengupta \emph{et~al.}, ``{Rethinking Split Manufacturing: An
  Information-theoretic Approach with Secure Layout Techniques},'' in
  \emph{{ICCAD}}.\hskip 1em plus 0.5em minus 0.4em\relax {IEEE}, 2017, pp.
  329--326.

\bibitem{MMSS:19}
T.~Moos \emph{et~al.}, ``{Glitch-Resistant Masking Revisited or Why Proofs in
  the Robust Probing Model are Needed},'' \emph{{IACR} TCHES}, vol. 2019,
  no.~2, pp. 256--292, 2019.

\bibitem{Pol:14}
I.~Polian, ``{Hardware Security and Test: Friends or Enemies?}'' \emph{it -
  Information Technology}, vol.~56, no.~4, pp. 192--202, 2014.

\bibitem{VSNFR:19}
E.~Valea \emph{et~al.}, ``{A Survey on Security Threats and Countermeasures in
  IEEE Test Standards},'' \emph{{IEEE} Des. {\&} Test}, vol. 36-3, pp. 95--116,
  2019.

\bibitem{PAA+:08}
I.~Parulkar \emph{et~al.}, ``{DFX} of a 3\({}^{\mbox{rd}}\) generation,
  16-core/32-thread ultrasparc- {CMT} microprocessor,'' in \emph{{ITC}}.\hskip
  1em plus 0.5em minus 0.4em\relax {IEEE}, 2008, pp. 1--10.

\bibitem{KGKP:18b}
B.~Karp \emph{et~al.}, ``{Detection and Correction of Malicious and Natural
  Faults in Cryptographic Modules},'' in \emph{PROOFS}, vol.~7, 2018, pp.
  68--82.

\bibitem{AARP:10}
J.~Aarestad \emph{et~al.}, ``{Detecting Trojans Through Leakage Current
  Analysis Using Multiple Supply Pad IDDQS},'' \emph{{IEEE} Trans. Information
  Forensics and Security}, vol.~5, no.~4, pp. 893--904, 2010.

\bibitem{RBIK:12}
F.~Regazzoni \emph{et~al.}, ``{Interaction Between Fault Attack Countermeasures
  and the Resistance Against Power Analysis Attacks},'' in \emph{Fault Analysis
  in Cryptography}, ser. Inf. Sec. and Crypt.\hskip 1em plus 0.5em minus
  0.4em\relax Springer, 2012, pp. 257--272.

\end{thebibliography}

\end{document}